\documentclass{pasj00}

\begin{document}
\SetRunningHead{I. Kayo et al.}{Three-Point Correlation Functions of SDSS Galaxies}
\Received{2004/3/14}%{yyyy/mm/dd}
\Accepted{2004/4/13}%{yyyy/mm/dd}

\title{Three-Point Correlation Functions of SDSS Galaxies 
in Redshift Space: Morphology, Color, and Luminosity Dependence}

%%% begin:list of authors
\author{%
Issha \textsc{Kayo},\altaffilmark{1}\thanks{Present address:  Department
of Physics and Astrophysics, Nagoya University, Nagoya 464-8602}
Yasushi \textsc{Suto},\altaffilmark{1}
Robert C. \textsc{Nichol},\altaffilmark{2}
Jun \textsc{Pan},\altaffilmark{3}
Istv\'an \textsc{Szapudi},\altaffilmark{3}\\
Andrew J. \textsc{Connolly},\altaffilmark{4}
Jeff \textsc{Gardner},\altaffilmark{4,5}
Bhuvnesh \textsc{Jain},\altaffilmark{6}
Gauri \textsc{Kulkarni},\altaffilmark{2}\\
Takahiko \textsc{Matsubara},\altaffilmark{7}
Ravi \textsc{Sheth},\altaffilmark{4}
Alexander S. \textsc{Szalay},\altaffilmark{8} and
Jon \textsc{Brinkmann}\altaffilmark{9}}
\altaffiltext{1}
{Department of Physics, School of Science, 
The University of Tokyo, Tokyo 113-0033} 
\altaffiltext{2}
{Department of Physics,  Carnegie
Mellon University, 5000 Forbes Avenue, Pittsburgh, PA 15213, USA}
\altaffiltext{3}
{Institute for Astronomy, University of Hawaii, 2680
Woodlawn Drive, Honolulu, HI 96822, USA}
\altaffiltext{4}
{Department of Physics and Astronomy, University of Pittsburgh, 3941 O'Hara Street, Pittsburgh, PA 15260, USA}
\altaffiltext{5}
{Pittsburgh Supercomputing Center, 4400 Fifth Ave., Pittsburgh, PA 15213,
USA}
\altaffiltext{6}
{Department of Physics, University of Pennsylvania,
Philadelphia, PA 19101, USA}
\altaffiltext{7}
{Department of Physics and Astrophysics, 
Nagoya University, Chikusa, Nagoya 161-8602}
\altaffiltext{8}
{Department of Physics and Astronomy, The Johns Hopkins University, \\
3701 San Martin Drive, Baltimore, MD 21218, USA}
\altaffiltext{9}
{Apache Point Observatory, Sunspot, NM 88349-0059, USA}

\email{kayo@a.phys.nagoya-u.ac.jp,  suto@phys.s.u-tokyo.ac.jp, 
nichol@cmu.edu, jpan@ifa.hawaii.edu,\\
 szapudi@ifa.hawaii.edu, 
ajc@phyast.pitt.edu, gardner@phyast.pitt.edu, bjain@hep.upenn.edu, \\
gkulkarn@andrew.cmu.edu, 
taka@a.phys.nagoya-u.ac.jp, sheth@sheth.phyast.pitt.edu, 
szalay@jhu.edu, jb@apo.nmsu.edu}
\maketitle

\begin{abstract}

We present measurements of the redshift--space three-point correlation
function of galaxies in the Sloan Digital Sky Survey (SDSS).  For the
first time, we analyze the dependence of this statistic on galaxy
morphology, color, and luminosity.  In order to control systematics due
to selection effects, we used $r$--band, volume-limited samples of
galaxies, constructed from magnitude-limited SDSS data
($14.5<r<17.5$), and further divided the samples into two morphological
types (early and late) or two color populations (red and blue).  The
three-point correlation function of SDSS galaxies follows the
hierarchical relation well, and the reduced three-point amplitudes in
redshift--space are almost scale-independent ($Q_z=0.5\sim1.0$).  In
addition, their dependence on the morphology, color, and luminosity is
not statistically significant.  Given the robust morphological, color
and luminosity dependences of the two-point correlation function, this
implies that galaxy biasing is complex on weakly non-linear to
non-linear scales. We show that a simple deterministic linear relation
with the underlying mass could not explain our measurements on these
scales.

\end{abstract}

\newpage
\section{Introduction}

While the first--year WMAP (Wilkinson Microwave Anisotropy Probe)
data imply that the primordial density fluctuations at $z\approx1000$
are virtually Gaussian \citep{K2003}, the present structure traced by
galaxies shows significant non-Gaussianity.  Such non-Gaussian features
naturally arise during the course of nonlinear
gravitational evolution of dark-matter density fields, the formation of
luminous galaxies and their subsequent evolution. Therefore, the degree
and nature of non-Gaussian signatures in the galaxy distribution will
provide important empirical constraints on the physics of galaxy formation.
The three-point correlation function (3PCF) is the lowest-order
unambiguous statistic to characterize such non-Gaussianities.

The determination of the 3PCF of galaxies was pioneered by
\citet{PG1975} using the Lick and Zwicky angular catalogs of galaxies.
\citet{GP1977} found that the 3PCF $\zeta (r_{12}, r_{23}, r_{31})$
obeys the following {\it hierarchical relation}:
%%%%%%%%%%%%%%%%%%%%%%%%%%%%%%%%%%%%%%%%%%%%%%%%%%%%%%%%%%%%%%%%%%
\begin{eqnarray}
&&  \zeta (r_{12}, r_{23}, r_{31}) \cr
&& = Q_{\rm r} \, [\xi(r_{12})\xi(r_{23}) + 
           \xi(r_{23})\xi(r_{31}) + \xi(r_{31})\xi(r_{12})],
\end{eqnarray}
%%%%%%%%%%%%%%%%%%%%%%%%%%%%%%%%%%%%%%%%%%%%%%%%%%%%%%%%%%%%%%%%%%
with $Q_{\rm r}$ being a constant, and $\xi(r)$ is the two-point correlation
function (2PCF). The value of $Q_{\rm r}$ in real space de-projected from
these angular catalogues is $1.29 \pm 0.21$ for $r \lesssim 3 h^{-1}{\rm
Mpc}$.  Although subsequent analyses of redshift catalogs confirmed the hierarchical
relation, at least approximately, the value of $Q_z$ (in redshift space)
appears to be smaller, $0.5\sim1$ (\cite{Bean1983}; \cite{Hale-Sutton};
Jing, B\"orner 1998). Recently, \citet{JB2004} have studied the
luminosity dependence of 
$Q$, and found a small, but significant, trend that brighter galaxies tend to
have a lower amplitude than fainter ones.

N-body simulations and perturbative analysis generally predict that the 3PCF
should depart from the hierarchical relation, especially on non-linear scales
(Fry 1984; Suto, Matsubara 1994; Jing, B\"{o}rner 1997; Barriga,
Gazta\~{n}aga 2002). However, Matsubara \& Suto (1994) demonstrated that
redshift distortions substantially reduce the scale-dependence of $Q_{\rm r}$,
resulting in $Q_z$ ($Q$ in redshift space) being nearly constant.  Still, the
amplitude of the $Q_z$ of galaxies is roughly (50\% $\sim$ 100\%) smaller than
predicted by N-body simulations.  The discrepancy is likely to be the effect
of biasing.  Recently, \citet{TJ2003} proposed a phenomenological model to
predict the 3PCF as a function of the galaxy properties, which is based on the
halo model.  They argue that the color dependence of $Q_{\rm r}$ should be
strong, i.e., the galaxy biasing severely affects the 3PCFs.

There are several statistics that are closely related to the 3PCF, e.g.,
the reduced moments of counts in cells, such as skewness and kurtosis (see
\cite{SzapudiEtal2002}, 2004 in preparation).  
Another similar statistics is the bispectrum, which is the Fourier
counterpart of the 3PCF. Recently, \citet{VerdeEtal2002} computed the
bispectrum of the 2dF Galaxy Redshift Survey, and concluded that the
non-linear bias is consistent with zero. The other complementary approach to
quantify the non-Gaussianity in the SDSS galaxies by means of topological
analysis (i.e., in terms of the Minkowski functionals) was already conducted
(Hoyle et al. 2002; Hikage et al. 2002, 2003). The above results are
largely consistent with
that predicted from purely gravitational nonlinearity, while a weak
morphological difference is marginally detected (e.g., figure 13 of
\cite{H2003}).  Here, we report on the first results of characterizing the
non-Gaussianity in the SDSS galaxy samples using 3PCFs in redshift space. In
particular, we consider the dependence of 3PCFs on the morphology, color, and
luminosity of galaxies by constructing volume-limited samples so as to avoid
possible systematics due to the selection function.  A separate companion
paper by Nichol et al. (in preparation) focuses on a detailed comparison
of 3PCFs between the 2dF (Jing, B\"orner 2004) and the SDSS galaxies.

The paper is organized as follows: section 2 describes our
volume-limited sample of SDSS galaxies and the morphology and color
classification methods. The measurements of 3PCFs are detailed in
section 3 with particular attention given to their dependence on the
morphology, color and luminosity of galaxies. Finally, we summarize the
results in section 4.  We compare several different estimators of 2PCFs
and 3PCFs in Appendix. Throughout the data analysis, we adopt the
following set of cosmological parameters: the matter density parameter,
$\Omega_{\rm m}=0.3$; the dimensionless cosmological constant,
$\Omega_\Lambda=0.7$; and the Hubble constant in units of 100km/s/Mpc,
$h=0.7$.

\section{Volume-Limited Samples of SDSS Galaxies}

It is expected that galaxy clustering depends on the intrinsic properties of
the galaxy samples under consideration, including their morphological types,
colors, and luminosities. Therefore, a straightforward analysis of
magnitude-limited samples must be interpreted with caution because several
different effects may be simultaneously involved. The magnitude-limited sample
of SDSS galaxies enables us to construct volume-limited samples of different
luminosities, morphologies and colors. This essentially removes the above
difficulties while keeping a statistically significant number of galaxies in
each sample.

Our present analysis is based on a subset of the SDSS galaxy redshift
data, `Large--scale Structure Sample 12' (Blanton et al. in preparation), which is
larger by a factor of 1.8 than the public Data Release One
\citep{Abazajian2003}. This sample
includes galaxies with $r$-band magnitudes of between $14.5$ and $17.77$
after correction for Galactic reddening using the maps of \citet{SFD98}.
For more details, the following can be consulted: \citet{Y2000}
for an overview of the SDSS;  
\citet{G1998} for a description of the photometric camera; 
 \citet{Stoughton2002} for photometric analysis;
\citet{Fukugita1996}, \citet{Hogg2001}, and \citet{Smith2002} for the
photometric system;
 \citet{Pier2003} for the astrometric calibration;
 \citet{E2001, Strauss2002} for  selection of the galaxy
spectroscopic samples; 
and \citet{B2003} for spectroscopic tiling.

%%%%%%%%%%%%%%%%%%%%%%%%%%%%%%%%%%%%%%%%%%%%%%%%%%%%%%%%%%%%%%%%%%%%%%
\begin{figure}[tbh]
  \centering \FigureFile(80mm,80mm){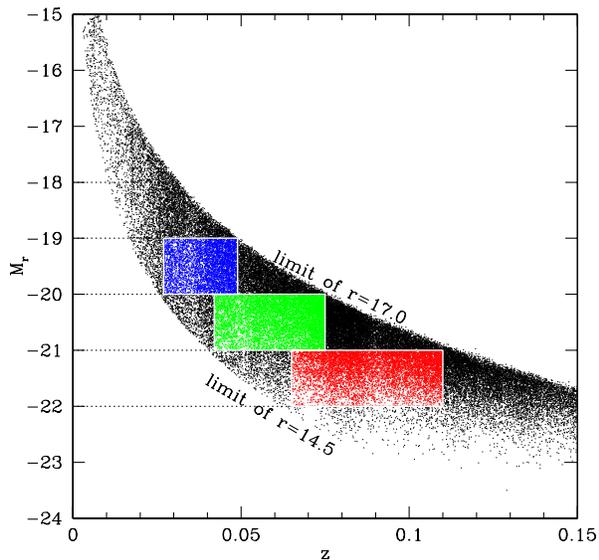}
\caption{Redshift - absolute magnitude relation for 
volume-limited samples of galaxies with $14.5<r<17.0$.
\label{fig:volume_limit_zM}}
\end{figure}
%%%%%%%%%%%%%%%%%%%%%%%%%%%%%%%%%%%%%%%%%%%%%%%%%%%%%%%%%%%%%%%%%%%%%%
%%%%%%%%%%%%%%%%%%%%%%%%%%%%%%%%%%%%%%%%%%%%%%%%%%%%%%%%%%%%%%%%%%%%%%
\begin{figure}[tbh]
  \centering \FigureFile(80mm,80mm){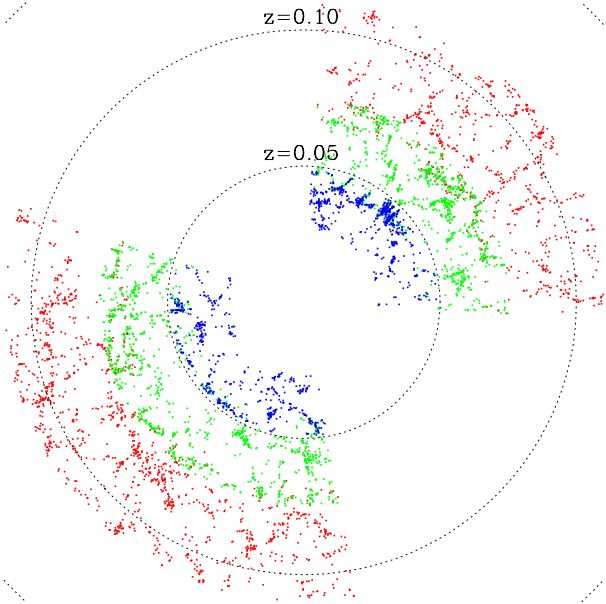}
\caption{Distribution of galaxies in the volume-limited samples
($14.5<r<17.0$; see table \ref{tbl:vm1} and figure
\ref{fig:volume_limit_zM}).  \label{fig:volume_limit_wedge}}
\end{figure}
%%%%%%%%%%%%%%%%%%%%%%%%%%%%%%%%%%%%%%%%%%%%%%%%%%%%%%%%%%%%%%%%%%%%%%

We construct two different kinds of volume-limited samples: one is based
on the morphological classification, and the other is on the color of
galaxies (figure \ref{fig:volume_limit_zM} and table \ref{tbl:vm1}).
Figure \ref{fig:volume_limit_wedge} illustrates the distribution of
those galaxies in a volume-limited sample.

%%%%%%%%%%%%%%%%%%%%%%%%%%%%%%%%%%%%%%%%%%%%%%%%%%%%%%%%%%%%%%%%
\begin{table}[thb]
 \caption{Volume-limited samples of SDSS galaxies; $14.5<r<17.0$ for the
 early/late classification and $14.5<r<17.5$ for color classification.
 \label{tbl:vm1}}
 \begin{center}
  \begin{tabular}{ccrrc}
   \hline\hline
   $M_r -5\log h$ & $z$ & early & late \\
   \hline
   $-22$ to $-21$ & 0.065-0.11 & 5881 & 3897  \\
   $-21$ to $-20$ & 0.042-0.075 & 5115 & 5975 \\
   $-20$ to $-19$ & 0.027-0.049 & 1626 & 3965 \\
   \hline \\
  \end{tabular}

  \begin{tabular}{ccrrc}
   \hline\hline
   $M_r -5\log h$ & $z$ & red & blue \\
   \hline
   $-22$ to $-21$ & 0.065-0.14 & 7949 & 8329  \\
   $-21$ to $-20$ & 0.042-0.093 & 8930 & 8155 \\
   $-20$ to $-19$ & 0.027-0.061 & 3706 & 3829 \\
   \hline
  \end{tabular}
 \end{center}
\end{table}
%%%%%%%%%%%%%%%%%%%%%%%%%%%%%%%%%%%%%%%%%%%%%%%%%%%%%%%%%%%%%%%%

\begin{figure}[tbh]
  \centering \FigureFile(80mm,80mm){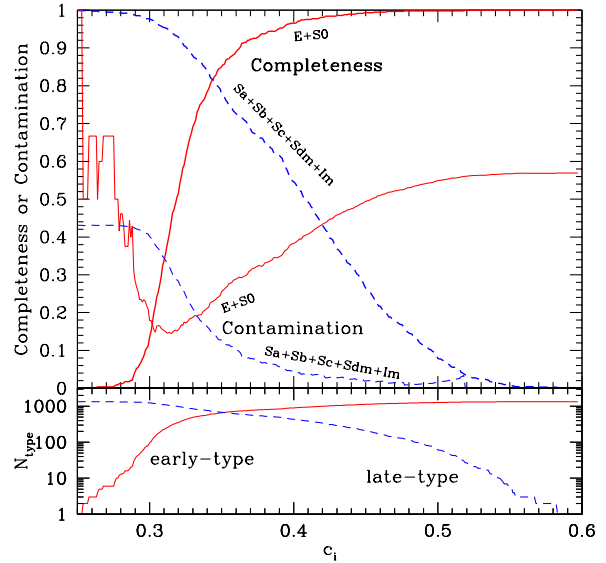}
\caption{Completeness of morphological classification according to the
 inverse concentration parameter.
\label{fig:morphology_cin}}
\end{figure}

Following Shimasaku et al.(2001), we classify the morphology of galaxies
according to the (inverse) concentration index ($c_i$), which denotes the ratio
of the half-light Petrosian radius to the 90\%-light Petrosian
radius. Specifically, we define galaxies with $c_i\le c_{i,c}$ as early-types,
and those with $c_i\ge c_{i,c}$ as late-types.  The critical value $c_{i,c}=
0.35$ is adopted for galaxies with $r<16.0$.  Figure \ref{fig:morphology_cin}
shows the completeness and contamination of this classification scheme. Around
$c_i\sim 0.35$ the completeness reaches 80\% and the contamination goes below
20\%.  The classification based on $c_i$ is affected by the fact that
the image quality of SDSS galaxies with $r>16.0$ starts to be degraded due to
the seeing. In order to empirically compensate for the effect, we slightly
change the critical value: $c_{i,c}=0.359$ for $16.0<r\le 16.5$ and
$c_{i,c}=0.372$ for $16.5<r\le 17.0$. Those values are chosen so that the
number ratio of early- to late-types remains the same in the three different
magnitude ranges: $14.0<r<16.0$, $16.0<r\le 16.5$, and $16.5<r\le 17.0$. We do
not classify the morphology of galaxies with $r>17.0$.

%%%%%%%%%%%%%%%%%%%%%%%%%%%%%%%%%%%%%%%%%%%%%%%%%%%%%%%%%%%%%%%%%%%%%%
%%%%%%%%%%%%%%%%%%%%%%%%%%%%%%%%%%%%%%%%%%%%%%%%%%%%%%%%%%%%%%%%%%%%%%
%%%%%%%%%%%%%%%%%%%%%%%%%%%%%%%%%%%%%%%%%%%%%%%%%%%%%%%%%%%%%%%%%%%%%%
\begin{figure}[tbh]
  \centering \FigureFile(80mm,80mm){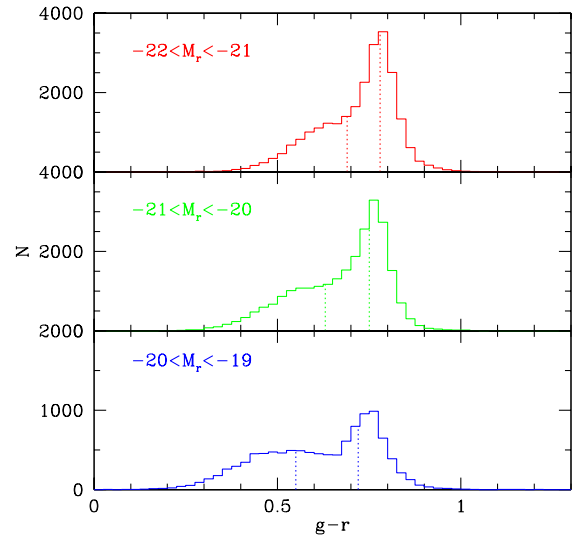}
 \caption{Histgrams of the
 color distribution for three volume-limited galaxy samples (table
 \ref{tbl:vm1}).  The bin of histogram is $\Delta(g-r) = 0.025$. The dotted
 vertical lines in each panel indicate the color thresholds that divide
 galaxies in three different color subsamples.  \label{fig:color}}
\end{figure}
%%%%%%%%%%%%%%%%%%%%%%%%%%%%%%%%%%%%%%%%%%%%%%%%%%%%%%%%%%%%%%%%%%%%%%

\begin{figure}[tbh]
  \centering \FigureFile(80mm,80mm){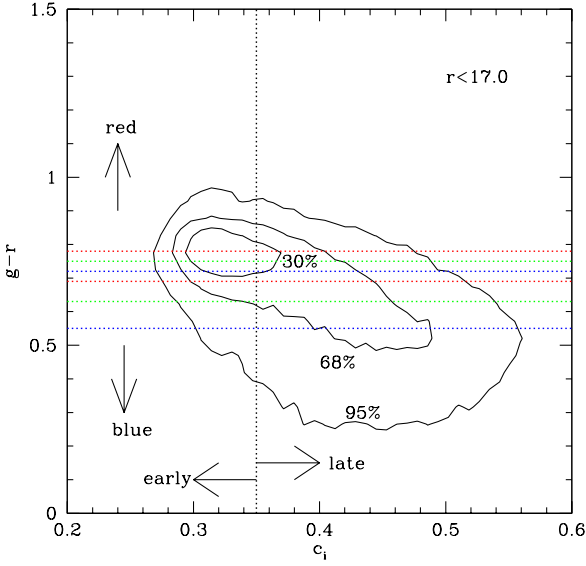}
\caption{Relation between the colors and the
 inverse concentration parameters of galaxies.
 The horizontal dotted lines correspond to the vertical dotted lines
 in figure~\ref{fig:color}.
\label{fig:color_cin}}
\end{figure}

On the other hand, classification of galaxies according to their colors
does not require such a good image quality as the $c_i$
classification. Thus, we can make use of galaxies down to $r=17.5$,
leading to an improved statistical significance.  Figure~\ref{fig:color}
shows the distribution of $g-r$ for volume-limited galaxies
corresponding to the lower part of table \ref{tbl:vm1}.  Clearly, the colors
are strongly correlated with the luminosity, and several different versions
of color-selected samples are possible. Since our goal is to see if
galaxy clustering is dependent on the colors in a complementary manner to
the previous morphological classification, we divided each
volume-limited sample into three subsamples that roughly have equal
numbers of
galaxies.  More precisely, we defined red (blue) galaxies as $g-r
>0.78$, $>0.75$, and $>0.72$ ($g-r <0.69$, $<0.63$, and $<0.55$) for
$-22<M_r-5\log h<-21$, $-21<M_r-5\log h<-20$, and $-20<M_r-5\log h<-19$,
respectively. Here, $g$ and $r$ are k-corrected magnitudes. These
thresholds are indicated in the vertical lines in figure
\ref{fig:color}.  Figure \ref{fig:color_cin} illustrates the relations
between our two classification schemes.

%%%%%%%%%%%%%%%%%%%%%%%%%%%%%%%%%%%%%%%%%%%%%%%%%%%%%%%%%%%%%%%%%%%%%%
%%%%%%%%%%%%%%%%%%%%%%%%%%%%%%%%%%%%%%%%%%%%%%%%%%%%%%%%%%%%%%%%%%%%%%

%%%%%%%%%%%%%%%%%%%%%%%%%%%%%%%%%%%%%%%%%%%%%%%%%%%%%%%%%%%%%%%%%%%%%%
\begin{figure}[tbh]
  \centering \FigureFile(80mm,80mm){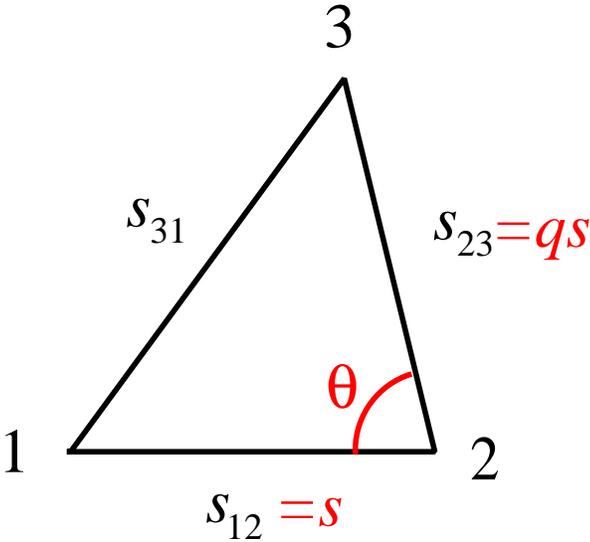}
\caption{Parameters to define the shape of triangles.
\label{fig:triangle}}
\end{figure}
%%%%%%%%%%%%%%%%%%%%%%%%%%%%%%%%%%%%%%%%%%%%%%%%%%%%%%%%%%%%%%%%%%%%%%

\section{Measurements and Results}

\subsection{Counting Triplets}

In the present analysis, we directly computed the number of triplets to
estimate the 3PCFs $\zeta(s_{12}, s_{23}, s_{31})$ in redshift space.
In this way, we properly take account of the complicated survey volume
boundary.  To be more specific, we distributed $N_{\rm R}$ random
particles over the survey volume of a given sample of $N_{\rm D}$
galaxies, following the selection function of the latter (we used $N_{\rm
R}=25N_{\rm D, all}$, where $N_{\rm D, all}$ is the number of all
galaxies in each luminosity bin).  Then, the 3PCFs were computed from the
following estimator (Szapudi, Szalay 1998):
%%%%%%%%%%%%%%%%%%%%%%%%%%%%%%%%%%%%%%%%%%%%%%%%%%%%%%%%%%%%%%%%%%%%%%
\begin{eqnarray}
&& \zeta(s_{12}, s_{23}, s_{31}) \cr
&& ~~~ =  \frac{DDD(s_{12}, s_{23}, s_{31})}{RRR(s_{12}, s_{23}, s_{31})}
-3\frac{DDR(s_{12}, s_{23}, s_{31})}{RRR(s_{12}, s_{23}, s_{31})} \cr
&& ~~~ +3\frac{DRR(s_{12}, s_{23}, s_{31})}{RRR(s_{12}, s_{23}, s_{31})}
-1,
\end{eqnarray}
%%%%%%%%%%%%%%%%%%%%%%%%%%%%%%%%%%%%%%%%%%%%%%%%%%%%%%%%%%%%%%%%%%%%%%%
where $DDD$, $DDR$, $DRR$, and $RRR$ are the number of corresponding
triplets consisting of galaxies and random particles, and are normalized
by $N_D(N_D-1)(N_D-2)/6$, $N_D(N_D-1)N_R/2$, $N_DN_R(N_R-1)/2$, and
$N_R(N_R-1)(N_R-2)/6$, respectively.  The above estimator is close to
optimal, because it is constructed to provide accurate edge effect
corrections.  We decided to use the Szapudi and Szalay estimator, since we
found that it converges the true 3PCF slightly more rapidly than the
estimator of \citet{JB1998} with the given number of random particles
(see Appendix for details).  We also estimated the redshift space 2PCF
with the analogous estimator (Landy, Szalay 1993), and defined the normalized
amplitude of 3PCFs as
%%%%%%%%%%%%%%%%%%%%%%%%%%%%%%%%%%%%%%%%%%%%%%%%%%%%%%%%%%%%%%%%%%%%%%%%%%
\begin{eqnarray}
&& Q_z(s_{12}, s_{23}, s_{31}) \cr
&& \equiv  \frac{\zeta(s_{12}, s_{23}, s_{31})}
  {\xi(s_{12})\xi(s_{23})+\xi(s_{23})\xi(s_{31})+\xi(s_{31})\xi(s_{12})}.
\label{eq:Qdef}
\end{eqnarray}
%%%%%%%%%%%%%%%%%%%%%%%%%%%%%%%%%%%%%%%%%%%%%%%%%%%%%%%%%%%%%%%%%%%%%%%%%%

Two different parameterizations of triangular shape are conventionally
used in this field. One is
%%%%%%%%%%%%%%%%%%%%%%%%%%%%%%%%%%%%%%%%%%%%%%%%%%%%%%%%%%%%%%%%%%%%%%%
\begin{eqnarray}
 s_{12}=s, ~ s_{23}= us , ~ s_{31}= (u+v)s ,
 \end{eqnarray}
%%%%%%%%%%%%%%%%%%%%%%%%%%%%%%%%%%%%%%%%%%%%%%%%%%%%%%%%%%%%%%%%%%%%%%
where $s_{12} \le s_{23} \le s_{31}$ is assumed (thus $u\geq 1$ and
$0\leq v < 1$). The other is 
%%%%%%%%%%%%%%%%%%%%%%%%%%%%%%%%%%%%%%%%%%%%%%%%%%%%%%%%%%%%%%%%%%%%%%%
\begin{eqnarray}
 s_{12}=s, ~ s_{23}=qs, ~ s_{31}= s\sqrt{1+q^2 -2q \cos\theta} ,
 \end{eqnarray}
%%%%%%%%%%%%%%%%%%%%%%%%%%%%%%%%%%%%%%%%%%%%%%%%%%%%%%%%%%%%%%%%%%%%%%
where $s_{12} \le s_{23}$ is assumed (thus, $q\geq 1$, but $s_{31}$ may be
smaller than either of the other two; see figure \ref{fig:triangle}). 
Note that according to the second
parameterization, one triangle is counted in three different bins in $(s,
q, \theta)$.  Although the first parameterization is more traditional in
the real data analysis, we adopt the second, which is widely used in
theoretical predictions. More specifically, we choose 8 equally-spaced
logarithmic bins in $0.4 h^{-1}{\rm Mpc} \leq s < 10.0 h^{-1}{\rm Mpc}$
and 5 equally-spaced linear bins both in $1\leq q < 5$ and in
$0<\theta<\pi$.  Since the number of triplets of $s< 1.0h^{-1}{\rm Mpc}$
is very small (typically $DDD\lesssim 50$), we plot the 3PCFs only for
$s>1.0h^{-1}{\rm Mpc}$ (where we have more than 100 $DDD$ counts) in
what follows. 
Also, the errors quoted below are evaluated from the 16
jack-knife re-sampling (see e.g., \cite{Lupton1993}).

\subsection{Equilateral Triangles}

\begin{figure}[tbh]
  \centering \FigureFile(80mm,80mm){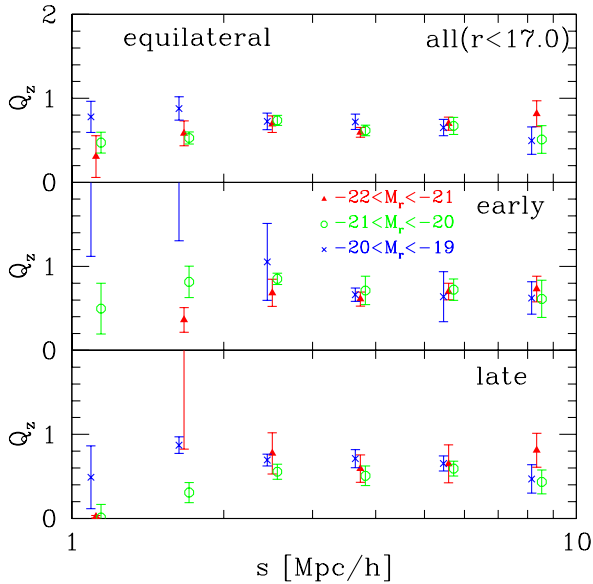}
\caption{Luminosity dependence of the normalized amplitude of three-point
correlation functions classified according to the morphologies of galaxies.
Only equilateral triangles are considered. Different symbols correspond
to different luminosities: $-22<M_r -5\log h< -21$ by solid
triangles, $-21<M_r -5\log h< -20$ by open circles, and $-20<M_r
-5\log h< -19$ by crosses.  Top: all galaxies, Middle:
early-type galaxies, Bottom: late-type galaxies.
\label{fig:Q_eq_cin_lum}}
\end{figure}

\begin{figure}[tbh]
  \centering \FigureFile(80mm,80mm){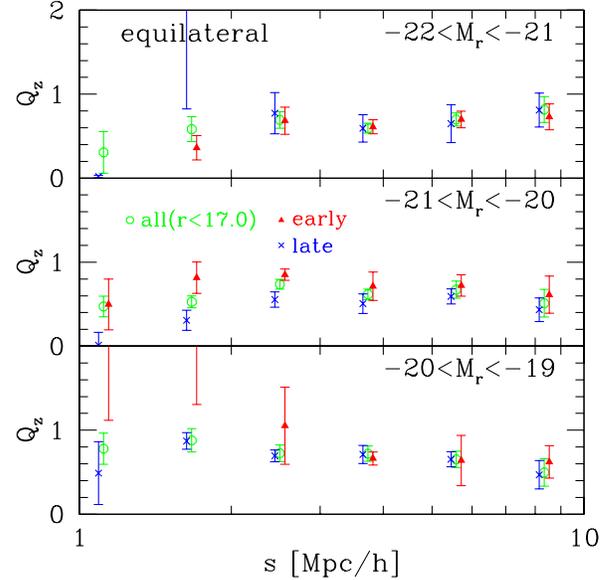} 
\caption{Same as figure \ref{fig:Q_eq_cin_lum}, but different symbols
corresponds to different morphologies; all galaxies in open circles,
early-type galaxies in solid triangles, and late-type galaxies in crosses.
Top: $-22<M_r -5\log h< -21$,
Middle: $-21<M_r -5\log h< -20$,
Bottom: $-20<M_r -5\log h< -19$. 
\label{fig:Q_eq_cin_mor}}
\end{figure}

\begin{figure}[tbh]
  \centering \FigureFile(80mm,80mm){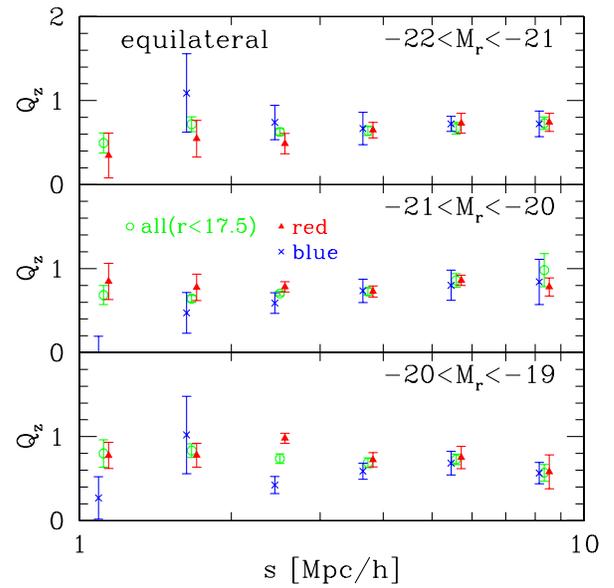}
\caption{Same as figure \ref{fig:Q_eq_cin_mor}, but
for classification according to the colors of galaxies. 
All galaxies by open circles,
red galaxies by solid triangles, and blue galaxies by crosses.
\label{fig:Q_eq_color_mor}}
\end{figure}

We ignore any possible dependence on the triangular shape for the
moment, and first consider 3PCFs for equilateral triangles in detail.
For this purpose, we first directly evaluate the lengths of three sides
of triplets. If those sides fall in the same bin among 8 equally-spaced
logarithmic bins between $0.4 h^{-1}{\rm Mpc}$ and $10.0 h^{-1}{\rm
Mpc}$, we define the triplets as being equilateral triangles.  Figures
\ref{fig:Q_eq_cin_lum}, \ref{fig:Q_eq_cin_mor}, and
\ref{fig:Q_eq_color_mor} show $Q_z$ over $1h^{-1}{\rm Mpc} < s < 10
h^{-1}{\rm Mpc}$.  The overall conclusion is that $Q_z$ is almost
scale-independent and ranges between 0.5 and 1.0, and no systematic
dependence is noticeable on the luminosity, morphology, and color.

Previous simulations and theoretical models (\cite{Suto1993}; Matsubara,
Suto 1994, \cite{M1994}, Takada, Jain 2003) indicate that $Q$ decreases
with the scale in both real and
redshift spaces. Observationally, \citet{JB2004} suggested that $Q_z$ of
the 2dF galaxies also decreases as scale.  Neither trend is clear in
the results. This might be partly due to the different velocity dispersion
of the galaxies in the two samples.  \citet{JMB1991} reported that the
early spirals (Sa, Sab, Sb) have a significantly smaller $Q$ than the
other types, which is qualitatively consistent with the present results.

%%%%%%%%%%%%%%%%%%%%%%%%%%%%%%%%%%%%%%%%%%%%%%%%%%%%%%%%%%%%%%%%%%%%%%
%%%%%%%%%%%%%%%%%%%%%%%%%%%%%%%%%%%%%%%%%%%%%%%%%%%%%%%%%%%%%%%%%%%%%%
%%%%%%%%%%%%%%%%%%%%%%%%%%%%%%%%%%%%%%%%%%%%%%%%%%%%%%%%%%%%%%%%%%%%%%
%%%%%%%%%%%%%%%%%%%%%%%%%%%%%%%%%%%%%%%%%%%%%%%%%%%%%%%%%%%%%%%%%%%%%%

%%%%%%%%%%%%%%%%%%%%%%%%%%%%%%%%%%%%%%%%%%%%%%%%%%%%%%%%%%%%%%%%%%%%%%
%%%%%%%%%%%%%%%%%%%%%%%%%%%%%%%%%%%%%%%%%%%%%%%%%%%%%%%%%%%%%%%%%%%%%%

\subsection{Relation to Biasing of the Two-Point Correlation Function}

The statement that the values of $Q_z$ are insensitive to
the intrinsic properties of galaxies may sound somewhat misleading.
The reduced three-point correlation amplitude is
normalized by the square of the 2PCFs for the corresponding galaxies.
This implies that the (unreduced) 3PCF depends on the galaxy
properties, since 2PCFs are known to show a clear dependence on the luminosity,
color, and morphology of galaxies (e.g., \cite{LS2003} for a recent
review). In order to further demonstrate the expected dependence in the
current samples , we compute the biasing parameters estimated from the
2PCFs,
%%%%%%%%%%%%%%%%%%%%%%%%%%%%%%%%%%%%%%%%%%%%%%%%%%%%%%%%%%%%%%%%%%%%%%
\begin{eqnarray}
 b_{z, i}(s) \equiv 
\sqrt{\frac{\xi_{z,i}(s)}{\xi_{z,{\rm \Lambda CDM}}(s)}} ,
\end{eqnarray} 
%%%%%%%%%%%%%%%%%%%%%%%%%%%%%%%%%%%%%%%%%%%%%%%%%%%%%%%%%%%%%%%%%%%%%%
where the index $i$ runs over each sample of galaxies with different
morphologies, colors, and luminosities.  The predictions of the mass
2PCFs in redshift space, $\xi_{z,{\rm \Lambda CDM}}(s)$, in the
$\Lambda$ Cold Dark Matter model are computed by properly taking
account of the light-cone effect over the corresponding redshift range
(Hamana et al. 2001).  In doing so, we adopt the fluctuation amplitude
$\sigma_8=0.9$ and the one-dimensional peculiar pairwise velocity
dispersion of 800km/s in addition to our fiducial set of cosmological
parameters.

\begin{figure}[tbh]
  \centering \FigureFile(80mm,80mm){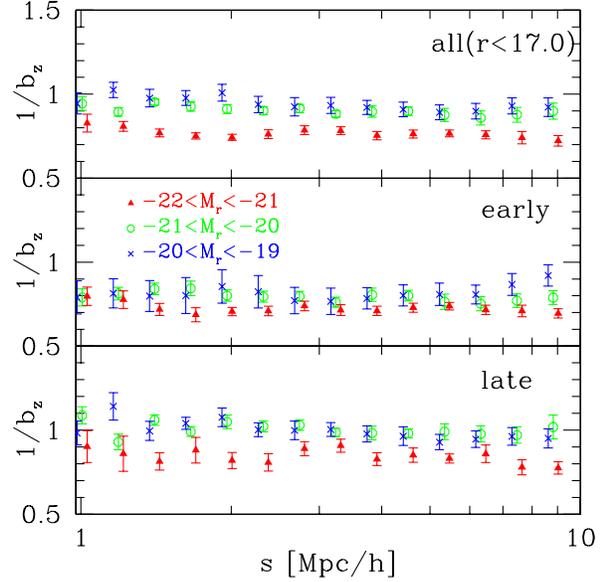}
\caption{Inverse of the biasing parameters of the two-point correlation
 functions plotted in the same way as in 
figure~\ref{fig:Q_eq_cin_lum}.
\label{fig:2ptbias_lum}}
\end{figure}

\begin{figure}[tbh]
  \centering \FigureFile(80mm,80mm){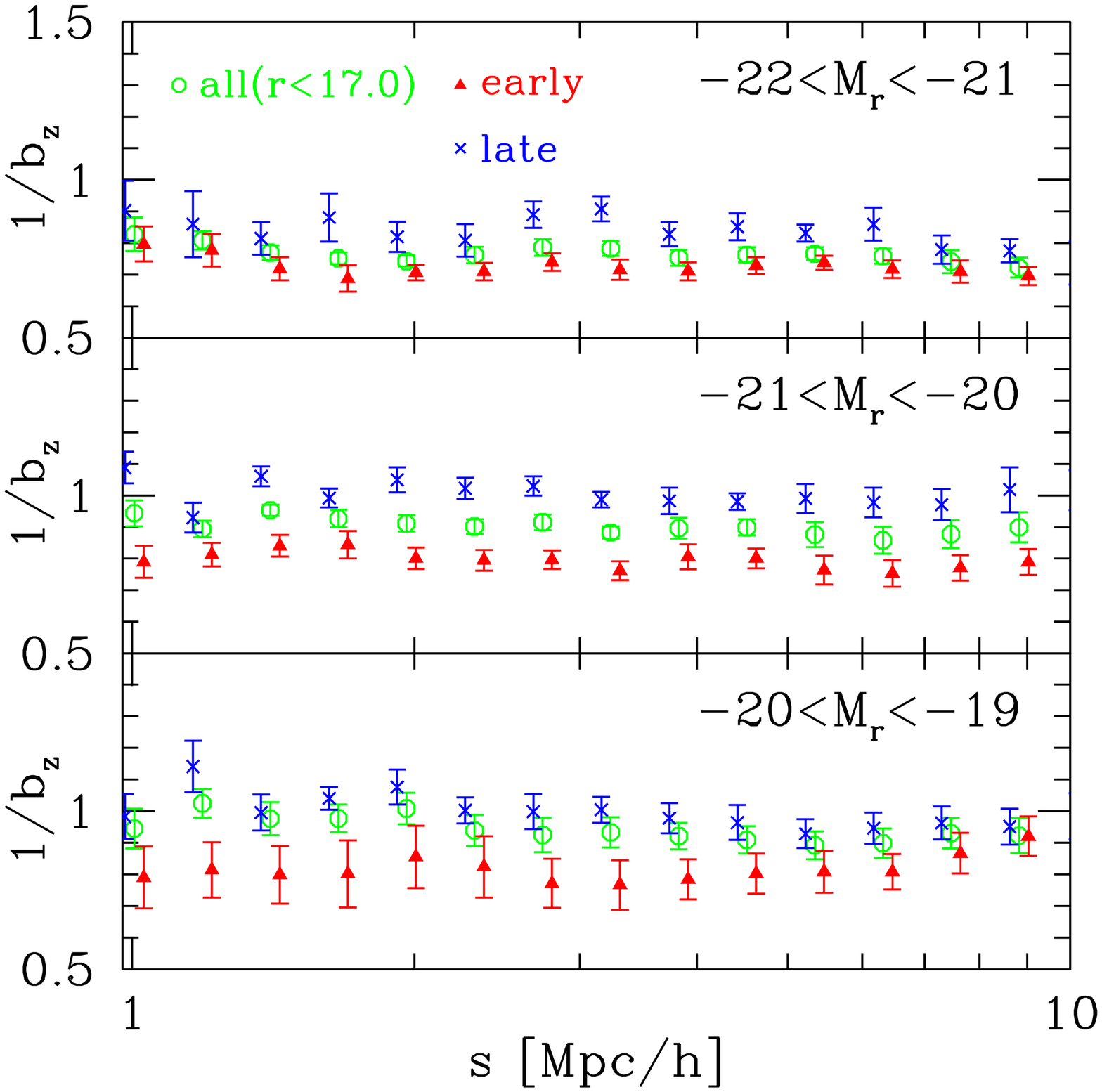}
\caption{Inverse of biasing parameters of the two-point correlation functions
 plotted in the same way as in figure~\ref{fig:Q_eq_cin_mor}.
\label{fig:2ptbias_mor}}
\end{figure}

\begin{figure}[tbh]
  \centering \FigureFile(80mm,80mm){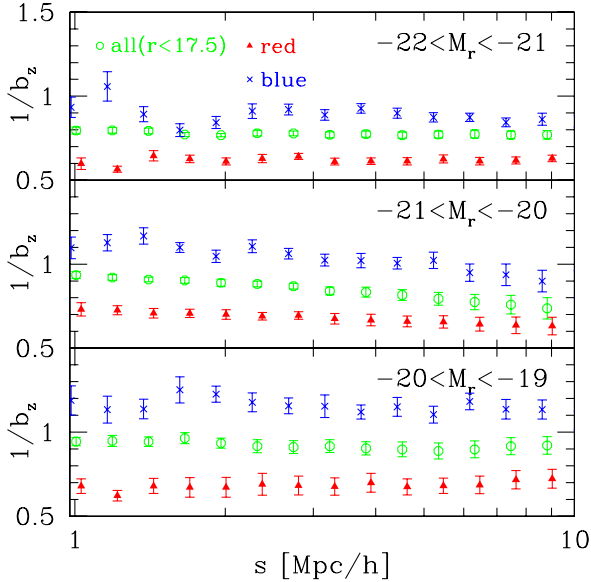}
\caption{Inverse of biasing parameters of the two-point correlation functions
 plotted in the same way as in figure~\ref{fig:Q_eq_color_mor}.
\label{fig:2ptbias_col}}
\end{figure}

As an illustrative example, consider a simple bias model in which the
galaxy density field, $\delta_{{\rm g}, i}$, for the $i$-th population of
galaxies is given by
%%%%%%%%%%%%%%%%%%%%%%%%%%%%%%%%%%%%%%%%%%%%%%%%%%%%%%%%%%%%%%%%%%%%%%
\begin{eqnarray}
\label{eq:quadratic_bias}
\delta_{{\rm g}, i} = b_{{\rm g}, i (1)}\delta_{\rm mass}
+ b_{{\rm g}, i (2)}\delta_{\rm mass}^2.
\end{eqnarray} 
%%%%%%%%%%%%%%%%%%%%%%%%%%%%%%%%%%%%%%%%%%%%%%%%%%%%%%%%%%%%%%%%%%%%%%
If both $b_{{\rm g}, i (1)}$ and $b_{{\rm g}, i (1)}$ are constant and the
mass density field $\delta_{\rm mass} \ll 1$, equation (\ref{eq:Qdef})
implies that
%%%%%%%%%%%%%%%%%%%%%%%%%%%%%%%%%%%%%%%%%%%%%%%%%%%%%%%%%%%%%%%%%%%%%%
\begin{eqnarray}
\label{eq:Q_quadratic_bias}
Q_{{\rm g}, i} = \frac{1}{b_{{\rm g}, i(1)}}
Q_{\rm mass} + \frac{b_{{\rm g}, i(2)}}{b_{{\rm g}, i(1)}^2} .
\end{eqnarray} 
%%%%%%%%%%%%%%%%%%%%%%%%%%%%%%%%%%%%%%%%%%%%%%%%%%%%%%%%%%%%%%%%%%%%%%
Thus, the linear bias model ($b_{{\rm g}, i(2)}=0$) simply implies that
$Q_{{\rm g}, i}$ is inversely proportional to $b_{{\rm g}, i(1)}$.
While this simple model may not be accurately applicable on the scales
of our results, it is instructive to plot $1/b_{{\rm g}, i}$ deduced
from the 2PCF of galaxies with different luminosities and
morphologies. The results are shown in figures \ref{fig:2ptbias_lum},
\ref{fig:2ptbias_mor}, and \ref{fig:2ptbias_col}.  A linear bias model
would predict that the ratio of $Q_z$ of early-types and late-types is
$Q_{\rm z, early}/Q_{\rm z, late} = b_{\rm late}/b_{\rm early} \approx
0.8$, and similarly that $Q_{\rm z, red}/Q_{\rm z, blue} = b_{\rm
blue}/b_{\rm red} \approx 0.5$.  Neither figure \ref{fig:Q_eq_cin_mor}
nor \ref{fig:Q_eq_color_mor}, however, shows such systematic trends
within our $\sim 20$ percent measurement accuracy. In a sense, the
biasing in the 3PCFs seems to compensate for the difference of $Q_{\rm g}$
purely due to that in the 2PCFs. Such a behavior is unlikely to be
explained by any simple model inspired by the perturbative expansion,
like equation (\ref{eq:quadratic_bias}). Rather, it indeed points to a
kind of regularity or universality of the clustering hierarchy behind
galaxy formation and evolution processes.  At least we can conclude that
the galaxy biasing is more complex than the simple deterministic and
linear model. More precise measurements of 3PCFs, and even higher order
statistics with future SDSS datasets would indeed be valuable to gain
more specific insights into the empirical biasing model.

%%%%%%%%%%%%%%%%%%%%%%%%%%%%%%%%%%%%%%%%%%%%%%%%%%%%%%%%%%%%%%%%%%%%%%
%%%%%%%%%%%%%%%%%%%%%%%%%%%%%%%%%%%%%%%%%%%%%%%%%%%%%%%%%%%%%%%%%%%%%%
%%%%%%%%%%%%%%%%%%%%%%%%%%%%%%%%%%%%%%%%%%%%%%%%%%%%%%%%%%%%%%%%%%%%%%
%%%%%%%%%%%%%%%%%%%%%%%%%%%%%%%%%%%%%%%%%%%%%%%%%%%%%%%%%%%%%%%%%%%%%%
%%%%%%%%%%%%%%%%%%%%%%%%%%%%%%%%%%%%%%%%%%%%%%%%%%%%%%%%%%%%%%%%%%%%%%
%%%%%%%%%%%%%%%%%%%%%%%%%%%%%%%%%%%%%%%%%%%%%%%%%%%%%%%%%%%%%%%%%%%%%%

\subsection{Shape Dependence}

\begin{figure}[tbh]
  \centering \FigureFile(80mm,80mm){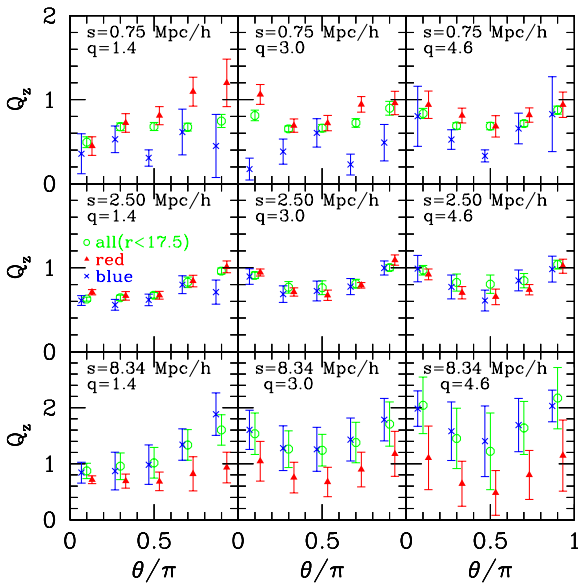}
\caption{Shape dependence of $Q_z$ of
volume-limited samples of galaxies 
 with $r<17.5$ and $-21<M_r -5\log h<-20$ 
(classified according to their color).
\label{fig:Qtheta_color}}
\end{figure}

Next consider the dependence of the 3PCFs on the triangular shape.  For
this purpose, we consider the volume-limited samples classified
according to the colors ($r<17.5$) because they have a greater number of
galaxies.  Figure \ref{fig:Qtheta_color} shows $Q_z$ for all, red, and
blue galaxies with $-21<M_r -5\log h<-20$ separately. Those plots
indicate a weak $\theta$-dependence expected from the previous
perturbation theory and N-body simulations (e.g., Barriga, Gazta\~naga
2002; Takada, Jain 2003).
The $\theta$-dependence is weaker on nonlinear scales ($s<1h^{-1}$Mpc),
and becomes noticeable on larger scales.  While the amplitudes of $Q_z$
are much smaller than, and the color dependence seems very different
from the corresponding prediction (figure 17 of Takada, Jain 2003), they
adopt a simple halo approach, and our color selection criteria are not
necessarily the same as theirs; further careful study is needed to
understand the origin of the difference (after the present paper is
accepted, \cite{WangEtal2004} submitted a
theoretical paper applying the halo model to account for the 3PCFs of
2dF galaxies, in particular their dependence on types and luminosity
dependences, and shows the possibility that the halo model can match
the observation).

%%%%%%%%%%%%%%%%%%%%%%%%%%%%%%%%%%%%%%%%%%%%%%%%%%%%%%%%%%%%%%%%%%%%%%
%%%%%%%%%%%%%%%%%%%%%%%%%%%%%%%%%%%%%%%%%%%%%%%%%%%%%%%%%%%%%%%%%%%%%%

\section{Summary and Discussion}

We have presented the first detailed study of the three-point
correlation function of SDSS galaxies in redshift space.  We
examined the dependence of their reduced amplitudes, $Q_z$,
[eq.(\ref{eq:Qdef})] on the scale and shape of the triangles for various
volumed-limited samples with different morphologies, colors, and
luminosities.  As for equilateral triangles, we basically confirm the
hierarchical clustering relation for scales of $1h^{-1}{\rm
Mpc}<s<10h^{-1}{\rm Mpc}$. On the other hand, $Q_z$ shows significant
shape dependence, as expected from perturbation theory and N-body
simulations, particularly at large scale, although the amplitude of $Q_z$
is much smaller than the expectation.  

The most important finding is that although the two-point correlation
functions of those galaxies show a clear dependence on their luminosity,
morphology, and color, we did not find any robust dependence of $Q_z$ on
those properties; the $Q_z$ results when taken alone imply that
galaxies are faithful tracers of the underlying mass distribution, and
that their intrinsic properties are independent of the purely nonlinear
gravitational evolution.  At the same time, however, their two-point
correlation functions strongly indicate that they are biased
tracers.  If we attempt to reconcile these two behaviors by using a
simple linear bias model, we cannot adjust it to simultaneously explain
both the two-point correlation functions and the insensitivity of $Q_z$
on the galaxy properties.  This implies that the galaxy biasing is
non-linear and fairly complex, while nonetheless displaying a remarkable
constancy in reduced amplitudes.

The above conclusion may appear to be inconsistent with the argument of
\citet{VerdeEtal2002}, who concluded that the non-linearity in 2dF
galaxy biasing is negligible from the bispectrum analysis. We note,
however, that they did not attempt to divide those galaxies into
different subsamples. If we look at the results of all galaxies alone,
top-panels of figures \ref{fig:Q_eq_cin_lum} and \ref{fig:2ptbias_lum}
may indicate the same conclusion as reached by \citet{VerdeEtal2002}.
In other words, our empirical finding indicates that the non-linearity
of the biasing, $b_2$, of galaxies vanishes when the corresponding linear
biasing parameter, $b_1$, is unity.  The fact that the two parameters,
$b_1$ and $b_2$, are highly correlated with each other is qualitatively
consistent with some biasing model predictions including halo models or
peak models (Mo et al. 1997).  It is remarkable that the current galaxy
survey approaches the level of discussing the validity of those models
quantitatively.  Future SDSS datasets with more area and a larger number
of galaxies will be able to place stronger limits on this
observation. This in turn will be useful for a detailed understanding of
the nature of galaxy biasing.

\bigskip
%%%%%%%%%%%%%%%%%%%%%%%%%%%%%%%%%%%%%%%%%%%%%%%%%%%%%%%%%%%%%%%%%%%%%%
We thank Y.P. Jing and Gerhard B\"orner for useful correspondences on their
measurements of three-point correlation functions of 2dF galaxies and an
anonymous referee for his/her useful comments.
I. K. gratefully acknowledges support from the Takenaka-Ikueikai fellowship.
He enjoyed the hospitality at Carnegie Mellon University and University of
Pittsburgh where part of this work was done, and thanks the Hayakawa fund of
Astronomical Society of Japan for the travel support.  Numerical computations
were carried out at ADAC (Astronomical Data Analysis Center) of the National
Astronomical Observatory of Japan, and also at computer
facilities at the University of Tokyo supported by the Special Coordination
Fund for Promoting Science and Technology, Ministry of Education, Culture,
Sports, Science and Technology. IS and JP was supported by NASA through AISR
NAG5-11996, and ATP NASA NAG5-12101 as well as by NSF grants AST02-06243 and
ITR 1120201-128440.

Funding for the creation and distribution of the SDSS Archive has been
provided by the Alfred P. Sloan Foundation, the Participating
Institutions, the National Aeronautics and Space Administration, the
National Science Foundation, the U.S. Department of Energy, the
Japanese Monbukagakusho, and the Max Planck Society. The SDSS Web site
is http://www.sdss.org/. 

The SDSS is managed by the Astrophysical Research Consortium (ARC) for
the Participating Institutions. The Participating Institutions are The
University of Chicago, Fermilab, the Institute for Advanced Study, the
Japan Participation Group, The Johns Hopkins University, Los Alamos
National Laboratory, the Max-Planck-Institute for Astronomy (MPIA),
the Max-Planck-Institute for Astrophysics (MPA), New Mexico State
University, University of Pittsburgh, Princeton University, the United
States Naval Observatory, and the University of Washington.

\bigskip

%%%%%%%%%%%%%%%%%%%%%%%%%%%%%%%%%%%%%%%%%%%%%%%%%%%%%%%%%%%%%%%%%%%%%%
\appendix
\section*{Comparison of the Estimators of Two- and
 Three-Point Correlation Functions}

For an accurate determination of the normalized amplitudes of $Q$, we
need reliable estimators for the two- and three-point correlation
functions.  For this purpose, we use a set of data particles
($N_D=64^3$; generated from a N-body simulation with $256^3$ dark matter
particles; \cite{JS1998}) distributed in a cube of $V_{\rm
box}=(100h^{-1} {\rm Mpc})^3$.  We then compare the following three
estimators for 2PCFs:
%%%%%%%%%%%%%%%%%%%%%%%%%%%%%%%%%%%%%%%%%%%%%%%%%%%%%%%%%%%%%%%%%%%%%%
\begin{eqnarray}
\label{eq:2ptdirect}
 \xi_{\rm direct} = \frac{DD}{V_{12}/V_{\rm box}} -1, &&~~~ ({\rm direct}); \\
\label{eq:2ptH}
 \xi_{\rm H} = \frac{DD\cdot RR}{[DR]^2} -1,  && ~~~ ({\rm Hamilton}~1993); \\
\label{eq:2ptLS}
 \xi_{\rm LS} = \frac{DD-2DR+ RR}{RR} , && ~~~ ({\rm Landy,~Szalay}~1993);
\end{eqnarray}
%%%%%%%%%%%%%%%%%%%%%%%%%%%%%%%%%%%%%%%%%%%%%%%%%%%%%%%%%%%%%%%%%%%%%%%
where $V_{12}$ denotes the volume of the spherical shell with the
thickness of the corresponding separation bin, and $DD$, $DR$, and $RR$
are the number of corresponding pairs of data and random particles, and
are normalized by $N_D(N_D-1)/2$, $N_D N_R$, and $N_R(N_R-1)/2$,
respectively.

Similarly, we use the following estimators for the 3PCFs:
%%%%%%%%%%%%%%%%%%%%%%%%%%%%%%%%%%%%%%%%%%%%%%%%%%%%%%%%%%%%%%%%%%%%%%
\begin{eqnarray}
\label{eq:3ptdirect}
\zeta_{\rm direct} 
=  \frac{DDD}{V_{123}^2/V_{\rm box}^2} - 3\xi_{\rm direct}-1, \\
\label{eq:3ptJB}
\zeta_{\rm JB} 
=  \frac{[RRR]^2 DDD}{[DRR]^3} -\frac{3RRR \cdot DDR}{[DRR]^2}+2 , \\
\label{eq:3ptSS}
\zeta_{\rm SS} 
 =  \frac{DDD-3DDR + 3DRR-RRR}{RRR}  .
\end{eqnarray}
%%%%%%%%%%%%%%%%%%%%%%%%%%%%%%%%%%%%%%%%%%%%%%%%%%%%%%%%%%%%%%%%%%%%%%%
The first uses the analytical expression for the effective volume
squared for the equilateral triplets. If one considers the size of the
the equilateral triplets between $r_{\rm min}$ and $r_{\rm max}$, it is
simply given by
%%%%%%%%%%%%%%%%%%%%%%%%%%%%%%%%%%%%%%%%%%%%%%%%%%%%%%%%%%%%%%%%%%%%%%
\begin{eqnarray}
V_{123}^2 = \pi^2 (r_{\rm max}^2-r_{\rm min}^2)^3 .
\end{eqnarray}
%%%%%%%%%%%%%%%%%%%%%%%%%%%%%%%%%%%%%%%%%%%%%%%%%%%%%%%%%%%%%%%%%%%%%%%
Because similar expressions in an arbitrary survey volume shape are
not available analytically, we focus on the equilateral triplets in
periodic cube of simulations.  The latter two correspond to estimators
by \citet{JB1998} and \citet{SS1998}, respectively.  The combination of
terms in \citet{JB1998} comes from the fact that the first term
corresponds to $1+\xi(r_{12})+\xi(r_{23})+\xi(r_{31})+\zeta(r_{12},
r_{23}, r_{31})$, and the second term to
$3+\xi(r_{12})+\xi(r_{23})+\xi(r_{31})$, yielding the desired $\zeta$
mathematically.

\begin{figure}[tbh]
 \centering \FigureFile(80mm,80mm){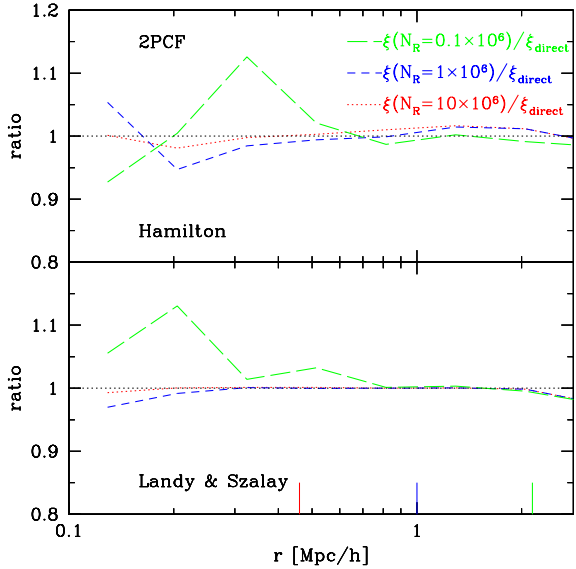}
 \caption{Comparison
of the three estimators of 2PCFs with a different number of random
particles ($N_R$). The small bars in the bottom panel indicate the mean
separation length of random particles for different $N_R$.
 Upper: the dependence of the Hamilton estimator on the number of random
particles. The long-dashed, short-dashed, and dotted curves denote
 $\xi_{\rm H}(N_R=0.1\times 10^6)/\xi_{\rm direct}$,
 $\xi_{\rm H}(N_R=1\times 10^6)/\xi_{\rm direct}$ and
 $\xi_{\rm H}(N_R=10\times 10^6)/\xi_{\rm direct}$, respectively.
 Lower: same as the upper panel, but for the Landy--Szalay estimator.
 \label{fig:convtest2p}}
\end{figure}

\begin{figure}[tbh]
  \centering \FigureFile(80mm,80mm){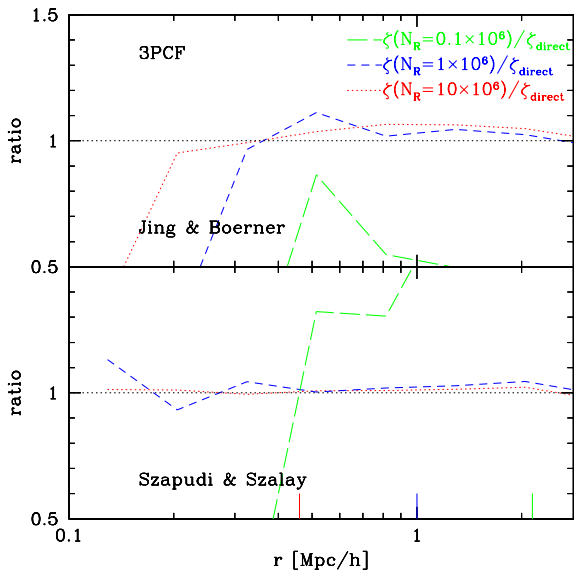}
\caption{Same as figure \ref{fig:convtest2p}, but for 3PCFs.
\label{fig:convtest3p}}
\end{figure}

The results of the convergence test are shown in Figure
\ref{fig:convtest2p} for 2PCFs and Figure \ref{fig:convtest3p} for
3PCFs.  These figures show the ratios of the two different estimators
with various $N_{\rm R}$ against the ``direct'' estimates
[equations (\ref{eq:2ptdirect}) and (\ref{eq:3ptdirect})] which do not use the
random particles.  With this number of random particles, equations
(\ref{eq:2ptLS}) and (\ref{eq:3ptSS}) reproduce the ``direct'' results
within $\sim 5\%$ for 2PCFs and 3PCFs, respectively.

Figure \ref{fig:convtest2p} indicates that $\xi_{\rm LS}$ converges to the
true value ($\xi_{\rm direct}$) more rapidly even with a smaller $N_R$
than $\xi_{\rm H}$ (see also Kerscher et al. 2000).  Figure
\ref{fig:convtest3p} shows that the estimators of 3PCFs are generally
less stable compared with those of 2PCFs. Still, $\zeta_{\rm SS}$ seems
better behaved than $\zeta_{\rm JB}$. Thus, we adopted $\xi_{\rm LS}$ and
$\zeta_{\rm SS}$ in the present analysis.

%%%%%%%%%%%%%%%%%%%%%%%%%%%%%%%%%%%%%%%%%%%%%%%%%%%%%%%%%%%%%%%%%%%%%%
%%%%%%%%%%%%%%%%%%%%%%%%%%%%%%%%%%%%%%%%%%%%%%%%%%%%%%%%%%%%%%%%%%%%%%

%%%%%%%%%%%%%%%%%%%%%%%%%%%%%%%%%%%%%%%%%%%%%%%%%%%%%%%%%%%%%%%%%%%%%%
%%%%%%%%%%%%%%%%%%%%%%%%%%%%%%%%%%%%%%%%%%%%%%%%%%%%%%%%%%%%%%%%%%%%%%

%%%%%%%%%%%%%%%%%%%%%%%%%%%%%%%%%%%%%%%%%%%%%%%%%%%%%%%%%%%%%%%%%%%%%%
%%%%%%%%%%%%%%%%%%%%%%%%%%%%%%%%%%%%%%%%%%%%%%%%%%%%%%%%%%%%%%%%%%%%%%

\end{document}